\documentstyle[twoside,fleqn,epsf,espcrc2]{article}

\newcommand{\AmS}{{\protect\the\textfont2
  A\kern-.1667em\lower.5ex\hbox{M}\kern-.125emS}}
\newcommand{\be}[1]{\begin{equation} \label{(#1)}}
\newcommand{\ee}{\end{equation}}
\newcommand{\ba}[1]{\begin{eqnarray} \label{(#1)}}
\newcommand{\ea}{\end{eqnarray}}

\def \Rpv{R_{P} \hspace{-1.2em}/\;\hspace{0.2em}}
\title
{Search for New Physics with Neutrinoless Double Beta Decay
\thanks{Talks presented by Heinrich P\"as at the 
Int. Europhysics Conference on 
High-Energy Physics, Jerusalem, August 19-26, 1997
and TAUP97, Gran Sasso, Sept. 7-11 1997}}

\author{
H.V. Klapdor--Kleingrothaus$^{\rm a}$
\thanks{Spokesman of the Collaboration},
L. Baudis\address{Max--Planck--Institut f\"ur Kernphysik, 
P.O. Box 103980,
D--69029 Heidelberg, Germany}, 
J. Hellmig$^{\rm a}$, M. Hirsch$^{\rm a}$,
S. Kolb$^{\rm a}$,
H. P\"as$^{\rm a}$
\thanks{present address: Laboratori Nazionali del Gran Sasso
(INFN), 67010 Assergi (AQ), Italy;
E-mail:
Heinrich.Paes@mpi-hd.mpg.de},\\  
Y. Ramachers$^{\rm a}$}

\begin{document}

\begin{abstract}
\noindent
Neutrinoless double beta decay ($0\nu\beta\beta$)
is one of the most sensitive approaches
to test particle physics beyond the standard model. 
During the last years, besides the most restrictive limit on the effective 
Majorana neutrino mass, the analysis of new 
contributions by the Heidelberg group led to bounds on 
left-right-symmetric models, leptoquarks and R-parity violating models 
competitive to recent accelerator limits, which are of special interest in 
view of the HERA anomaly at large $Q^2$ and $x$. 
These new results deduced from the Heidelberg-Moscow double 
beta decay experiment are reviewed. Also an outlook on the future of double 
beta decay, the GENIUS proposal, is given.
\end{abstract}

\maketitle

\section{Introduction}
Double beta decay \cite{Kla98,tren} corresponds to two single beta decays 
occuring in 
one
nucleus and 
converts a nucleus (Z,A) into a nucleus (Z+2,A).
While even the standard model allowed process emitting two antineutrinos
\be{i1}
^{A}_{Z}X \rightarrow ^A_{Z+2}X + 2 e^- + 2 {\overline \nu_e}
\ee
is one of the rarest processes in nature with half lives in the region of
$10^{21-24}$ years, more interesting is the search for 
the neutrinoless mode,
\be{i2}        
^{A}_{Z}X \rightarrow ^A_{Z+2}X + 2 e^- 
\ee
which
violates lepton number by two units and thus implies physics beyond the 
standard model.

\section{The Heidelberg--Moscow Double Beta Decay Experiment}
The Heidelberg--Moscow experiment \cite{HM97,HM97b,Kla98}
is a second generation experiment
searching for the $0\nu\beta\beta$ decay of $^{76}Ge$. 
Five crystals with an active mass of 10.96 kg, 
grown out of 19.2 kg of 86\% enriched $^{76}$Ge, are in regular operation
as p--type HPGe detectors in low level cryostats in the Gran Sasso laboratory,
which provides a shielding of 3500 m of water equivalent (mwe).
The high source strength of the experiment, the large size of the detectors 
concentrating the
background in the peaks and the excellent energy resolution yield an
outstanding position compared with other experiments.
It has been described recently in detail in \cite{HM97,HM97b,Kla98}.

Fig. 1 shows the results after 31.8 kg y
measuring time for all data corresponding to a half life limit of
\be{exp1}
T_{1/2}^{0\nu\beta\beta}>1.2 \cdot 10^{25} y.
\ee 
The limits from the pulse shape 
\cite{hel,HM97b}
data with 15.3 kg y (filled histogram in Fig. 
1), corresponding to
$T_{1/2}^{0\nu\beta\beta}>1.1 \cdot 10^{25} y$, are not yet competitive 
to the large data set without application of pulse shape analysis.
However the background improvement will allow to test the half life 
region up to $6 \cdot 10^{25}$ y, corresponding to a neutrino mass limit of 
0.1--0.2 eV, during the next five years.

The standard model allowed $2\nu\beta\beta$ decay is measured with the 
highest statistics ever reached, containing 21115 events in the energy region
of 500-2040 keV, and yields a half life of \cite{HM97}
\be{exp2}
T_{1/2}^{2\nu\beta\beta}=(1.77^{+0.01}_{-0.01}(stat.)^{+0.13}_{-0.11}(syst.)) 
\cdot 10^{21} y.
\ee
This result, confirming the theoretical predictions of 
\cite{staudt}
with an accuracy of
a factor $\sim \sqrt{2}$, provides a consistency check 
of nuclear matrix 
element calculations. 
It also for the first time opens up the possibility to 
search for deviations of the $2\nu\beta\beta$ spectrum such as those due to
emission of exotic scalars \cite{major}.

\begin{figure}[!t]
\vspace*{-20mm}
\hspace*{3mm}
\epsfxsize=70mm
\epsfbox{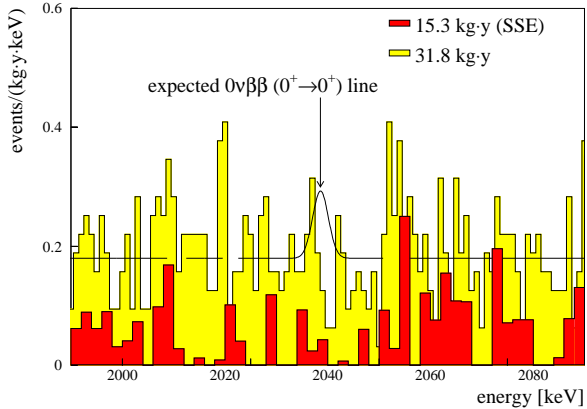}
\vspace*{-32mm}
\caption{Combined spectrum in the $0\nu\beta\beta$ peak region: 
Complete data with 31.8 kg $\cdot$y,
with pulse shape analysis detected single site events with 15.3 kg 
$\cdot$ y and
90\% C.L. excluded peak.
\label{15}}
\vspace*{-8mm}
\end{figure}

\section{Double Beta Decay and Physics Beyond the Standard Model}

\subsection{Neutrino Mass}
The search for $0\nu\beta\beta$ decay exchanging a massive 
left--handed Majorana
neutrino between two standard model vertices 
at present provides the most 
sensitive approach to determine the mass of the neutrino and also 
a unique possibility to distinguish between the Dirac or  
Majorana nature of the neutrino.
With the recent half life limit of the Heidelberg--Moscow
experiment \cite{HM97b,Kla98} the following limits on effective left--handed 
neutrino masses can be deduced:
\be{numass}
\langle m_{\nu}\rangle \leq 0.45 eV \hskip5mm (90\% C.L.)
\ee  
\be{numass2}
\langle m_{\nu}\rangle \geq 7.5\cdot 10^7 GeV \hskip5mm(90\% C.L.)
\ee
Taking into account the uncertainties in the numerical values of
nuclear matrix elements of about a
factor of 2, the Heidelberg--Moscow experiment, improving its half life limit
up to $6 \cdot 10^{25}y$, will test degenerate neutrino 
scenarios \cite{deg} in the next five years. 

\subsection{Left--Right--Symmetric Models}
In left--right symmetric models the left--handedness of weak 
interactions is explained as due to the effect of different symmetry breaking 
scales in the left-- and in the right--handed sector. 
$0\nu\beta\beta$ decay proceeds through exchange of the heavy right--handed 
partner of the ordinary neutrino between right-handed W vertices, leading
to a limit of
\be{mwr}
m_{WR}\geq 1.2 \Big(\frac{m_N}{1TeV}\Big)^{-(1/4)} TeV.
\ee 
Including a theoretical limit obtained from considerations of vacuum 
stability \cite{moha86} one can deduce an absolute lower limit on the 
right--handed W mass of \cite{11} 
\be{mwr2}
m_{W_{R}}\geq 1.2 TeV.
\ee

\subsection{Supersymmetry}
Supersymmetry (SUSY), providing a symmetry between fermions and bosons and
thus doubling the particle spectrum of the SM, belongs
to the most prominent extensions of the standard model. 
While in the minimal supersymmetric extension (MSSM) R--parity
is assumed to be conserved, there are no theoretical reasons for $R_p$ 
conservation and
several GUT \cite{rpguts} and Superstring \cite{rpss} models require
R--parity violation in the low energy regime.  
Also recent reports concerning an anomaly at HERA in the inelastic 
$e^+p$
scattering 
at high $Q^2$ and $x$ \cite{HERA} renewed the interest in $\Rpv$--SUSY
(see for example \cite{kalino}).
In this case $0\nu\beta\beta$ decay can occur through Feynman graphs 
involving the exchange of
superpartners as well as $\Rpv$--couplings $\lambda^{'}$ 
\cite{hir95,hir95d,hir96c,hir96}.
The half--life limit of the Heidelberg--Moscow experiment leads to bounds
in a multidimensional parameter space \cite{hir95,hir96c}
\be{susy5}
\lambda_{111}^{'}\leq 3.2\times 10^{-4}\Big(\frac{m_{\tilde{q}}}{100 
GeV} \Big)^2
\Big(\frac{m_{\tilde{g}}}{100 GeV} \Big)^{1/2}
\ee
(for $m_{\tilde{d}_{R}}=m_{\tilde{u}_{L}}$), which are the sharpest limits on
 $\Rpv$--SUSY.
This limit also excludes the first generation squark interpretation of the
HERA events \cite{Hir97b}.
In the case of R--parity conserving SUSY, based on a theorem proven in 
\cite{sneut},
the $0\nu\beta\beta$ mass limits can be converted in sneutrino Majorana mass
limits being more restrictive than what could be obtained
in inverse neutrinoless double beta 
decay and single sneutrino production at future linear colliders (NLC)
\cite{sneut}.
 
\subsection{Leptoquarks}
Leptoquarks are scalar or vector particles coupling both to leptons 
and quarks,
which appear naturally in GUT, extended Technicolor
or Compositeness models containing leptons and 
quarks in the same multiplet. 
Also production  of a scalar leptoquark with mass of $m_{LQ}\simeq 200GeV$ 
has been considered as a solution to the HERA 
anomaly (see for example \cite{kalino}). 
However, TEVATRON searches have set stringent limits of 
(combined with NLO cross section calculations \cite{kraem97})
$m_{LQ}>240GeV$ for scalar leptoquarks decaying with branching ratio 1
into electrons and quarks. 
One possibility   to keep the leptoquark
interpretation interesting is therefore to reduce the branching ratio due
to the mixing of different multiplets leading to a signifi\-cant weakening 
of the CDF/D0 limits \cite{babu97}. This kind of mixing can be obtained by 
introducing a leptoquark--Higgs coupling -- which would lead  to a contribution
to $0\nu\beta\beta$ decay
\cite{hir96a}. Combined with the half--life limit of the 
Heidelberg--Moscow experiment bounds on effective couplings can be derived
\cite{lepbb}. 
Assuming 
only one lepton number violating $\Delta L=2$ LQ--Higgs coupling unequal 
to zero and  the leptoquark masses 
not too different, one can derive from this limit either a bound on the 
LQ--Higgs coupling
\ba{erg4}
Y_{LQ-Higgs}=(few)\cdot 10^{-6}
\ea
or a limit implying that HERA does not see Leptoquarks with masses in the 
range of
${\cal O}(200 GeV)$. This excludes 
most of the possibilities to
relax the TEVATRON bounds by introducing LQ--Higgs couplings to reduce the
branching ratio \cite{Hir97b}.

\section{Outlook on the future of double beta decay: The Heidelberg
project GENIUS}
To render possible a further breakthrough in search for neutrino masses and 
physics beyond the standard model, recently GENIUS, an experiment operating
a large amount of naked Ge--detectors in a liquid nitrogen shielding,
has been proposed (\cite{Kla98} and studied in detail in \cite{gen}).
The possibility to operate Ge detectors inside liquid nitrogen has already been
demonstrated by the Heidelberg group and yield an excellent energy resolution 
and threshold.

Operating 288 enriched $^{76}$Ge detectors with a total mass of 1  ton 
inside a nitrogen tank of 9 m height and diameter, improves 
the sensitivity to neutrino masses down to 0.01 eV. This  allows to solve
the atmospheric neutrino problem, if it is due to 
$\nu_{e} \leftrightarrow \nu_{\mu}$ oscillations, confirm or exclude
Majorana 
neutrinos as hot dark matter in the universe as well as to test SUSY models, 
leptoquarks and
right--handed W--masses comparable to the LHC \cite{Kla98,gen}, requiring
purity levels less stringent (except for $^{222}$Rn) as already obtained by 
the CTF (Counting Test Facility) of the 
Borexino experiment.
A ten ton version would probe neutrino masses even down to $10^{-3}$ eV, 
which 
would allow to test the large angle MSW solution of the solar neutrino 
problem. 
As direct dark matter detection experiment it would allow to test almost 
the entire MSSM parameter space already in a first step using only 100 kg 
of enriched or even natural Ge \cite{Kla98,gen}.

\section{Conclusions}
Neutrinoless double beta decay has a broad potential to test physics 
beyond the standard model. The possibilities to constrain neutrino masses,
left--right--symmetric models, SUSY and leptoquark scenarios 
have been reviewed. Experimental 
limits on $0\nu\beta\beta$ decay are not only complementary to 
accelerator experiments but  at least in some cases competitive or superior
to the best existing direct search limits. The Heidelberg--Moscow experiment
has reached the leading position among double beta decay experiments and as
the first of them now yields results in the sub--eV range for the neutrino 
mass. A further breakthrough will be possible realizing the GENIUS proposal. 
For the application of double beta technology in WIMP dark matter
search we refer to \cite{baud}.

\end{document}